\documentclass[twocolumn,pre,aps,showpacs]{revtex4}
\usepackage[english]{babel}
\usepackage{fontenc}
\usepackage{amsmath}
\usepackage{amssymb}
\usepackage{epsfig}
\usepackage{graphicx}
\usepackage{dcolumn}
\usepackage{bm}
\usepackage{color}

\begin{document}

\title{Detection of trend changes in time series using Bayesian inference}

\author{N. Sch\"{u}tz}
\author{M. Holschneider}
\affiliation{Focus Area for Dynamics of Complex Systems, Universit\"{a}t Potsdam, Karl-Liebknecht-Str. 24,
D-14476 Potsdam, Germany}

\date{Received: \today / Accepted: date}

\begin{abstract}
Change points in time series are perceived as isolated singularities where two regular trends of a given signal do not match.
The detection of such transitions is of fundamental interest for the understanding of the system's internal dynamics.
In practice observational noise makes it difficult to detect such change points in time series.
In this work we elaborate a Bayesian method to estimate the location of the singularities and to produce some confidence intervals.
We validate the ability and sensitivity of our inference method by estimating change points of synthetic data sets. 
As an application we use our algorithm to analyze the annual flow volume of the Nile River at Aswan from 1871 to 1970, where 
we confirm a well-established significant transition point within the time series.
\end{abstract}

\maketitle


\section{Introduction}
\label{intro}

The estimation of change points challenges analysis methods and modeling concepts. Commonly change points are considered 
as isolated singularities in a regular background indicating the transition between two regimes governed 
by different internal dynamics.
In time series analysts focus on change points in observed data to reveal dynamical properties of the system 
under study and to infer on possible correlations between subsystems.
Detecting trend changes within various data sets is under intensive investigation in numerous research disciplines, such as 
palaeo-climatology \cite{Trauth_2009,Mudelsee_2005}, ecology \cite{Girardin_2009,Jong_1998}, 
bioinformatics \cite{Minin_2005, Liu_1999} and economics \cite{Lia_2007,Andrews_1993}.\\
In general, the detection of transition points is adressed via {\bf(i)} regression \cite{Mudelsee_2009} or 
{\bf(ii)} spectral analysis methods \cite{Olsen_2008}, {\bf(iii)} Bayesian approaches \cite{Moreno_2005,Lian_2009} or 
{\bf(iv)} recurrence network techniques \cite{Donner_2010,Marwan_2009}.\\
In this work we formulate transition points not only in terms of the underlying regular dynamics, but also
as a transition in the heteroscedastic noise level. We use Bayesian inference to produce estimates for all relevant 
parameters. 
Our signal model is described by a regular mean undergoing a sudden change and a heteroscedastic fluctuation which 
undergoes as well a sharp transition at the same time point. Thus, in its simplest form, the observed signal 
$\bm{y}$ has a linear trend undergoing a break point $\theta$ at a time point $t_{i} = \theta$. 
The posterior density $p(\theta|\bm{y})$ of the change point given the signal enables us to derive the point estimate 
$\hat{\theta}$ as the most likely break point and its confidence bounds.
By applying a sliding window, we formally localize the posterior density and the modelling of the subsignals as a linear trend 
is valid in first order. Consequently we investigate time series globally and locally for a generalized break point 
in the signal's statisitical properties.\\
In comparison to established methods (e.g. {\bf(ii)} multiscale spectral analysis \cite{Olsen_2008}) our technique is not restricted on 
a uniform time grid (e.g. as required for filtering methods). 
The majority of existing methods require additional approaches to interpret the confidence of the outcome (e.g. {\bf(i)} bootstrapping, 
{\bf(ii)} test statistics, {\bf(iv)} introducing measures). Whereas our technique provides the confidence 
intervals of the estimates as a byproduct in a natural way. This, for us is actually 
the most convincing argument to approach the detection task via Bayesian inference since
besides the parameter estimation on its own, we obtain a degree of belief about our assumed model 
and about the uncertainties in the parameters \cite{DAgostini_2003,BatesDebRoy_2004,Gelman}. 
Existing techniques addressing Bayesian inference {\bf(iii)} approach on the 
one hand the plain localization task of the singularity by treating the remaining model's parameter as hidden \cite{Fearnhead_2006,Downey_2008}. 
On the other hand hierarchical Bayesian models are used \cite{Moreno_2005} mainly based on Monte-Carlo-expectation-maximization 
(MEMC) algorithms for the estimation process \cite{Liu_1999,Lian_2009}.\\
In contrast, we intend to achieve an insight in the parameter structure of the time series. We intend to detect 
multiple change points without enlarging the model's dimensionality, since this increases considerably the computational time. 
By addressing the general framework of linear mixed models (LMM) \cite{McCulloch} we are able to 
factorize the joint posterior density into a family of parametrized Gaussians. This mirrors the
separation of the linear from the non-linear parts and it simplifies considerably the explicit computation of the marginal distributions.
Our technique will be applied to a hydrological time series of the river Nile, which exhibits a well known change point. 


\section{Definition of the model}

In our modeling approach we consider two aspects of change points in a time series. On the one hand, a change point is 
commonly associated with a sudden change of local trend in the data. This indicates a transition point between
two regimes governed by two different internal dynamics. 
On the other hand we assume that the systematic evolution of the local variability of the data around its average value 
undergoes a sudden transition at the change point. 
As we will show, both aspects can be combined into a linear mixed model with hyperparameters. 
Our formulation allows the separation of the Gaussian from the intrinsic non-linear parts of
the estimation problem, which besides clarifying the structure of the model, speeds up computations considerably.

\subsection{Formulation of the linear mixed model}
\label{lmm}

\begin{figure}
        \includegraphics[width=0.5\textwidth]{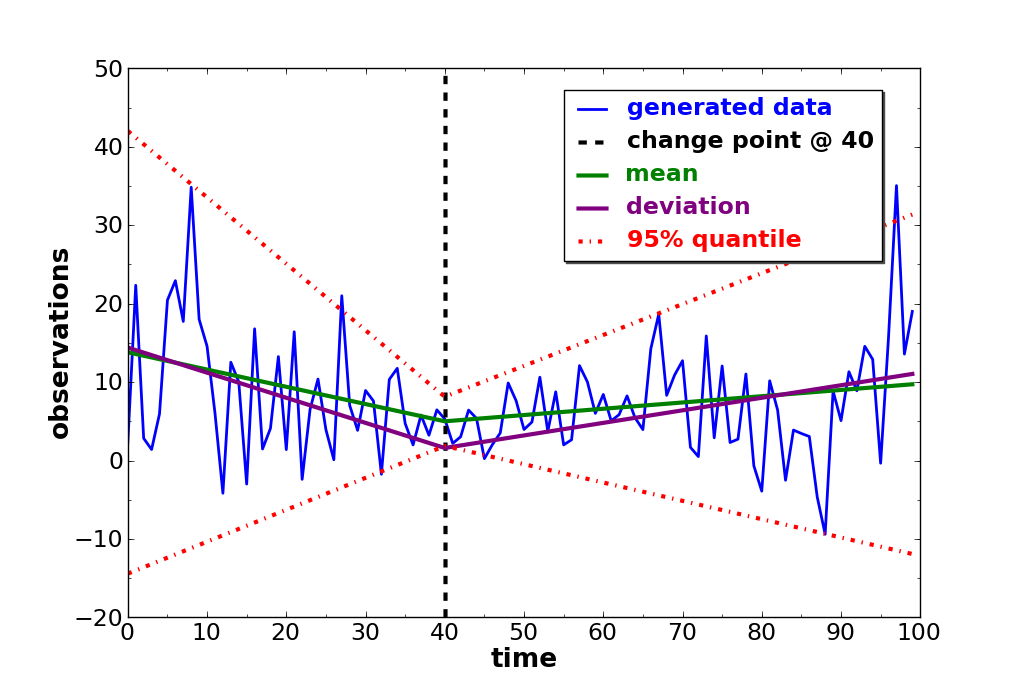}
	\caption{Realization of a synthetic time series of $n_{obs}=100$ data points generated by Equ.(\ref{lm}) whereas 
                 the mean is parametrized by
                 $F_{\theta}\bm{\beta}= 5 + 0.22\cdot\bm{\zeta}^{\theta}_{-} + 0.08\cdot\bm{\zeta}^{\theta}_{+}$ 
                 and the deviation is modeled as
                 $\sigma^{2} \Omega_{\theta, \bm{s}}= \left[1.6(1 + 0.2\cdot\bm{\zeta}^{\theta}_{-} + 0.1\cdot\bm{\zeta}^{\theta}_{+})\right]^{2}$.}
	\label{fig:1} 
    \end{figure}

The simplest type of signal undergoing a change point at time $\theta$ can be expressed as
    \begin{equation}
y(t) = \beta_{0} \,+\, \beta_{1} |\theta - t|_{-} \,+\, \beta_{2} |\theta - t|_{+} \,+\, \xi(t)\,.
    \end{equation}
Here we use the elementary Hockey sticks of first order defined through
    \begin{equation}
     |\theta-t|_{\tiny -}= \left(\zeta^{\theta}_{-}\right) = \begin{cases}& \theta-t \hspace{2mm} \text{if} \hspace{2mm} t \leq \theta \\& 0 \hspace{9mm}  \text{else} \hspace{8mm} \end{cases}, \label{hockey1} 
    \end{equation}
and
    \begin{equation}
     |\theta-t|_{\tiny +}= \left(\zeta^{\theta}_{+}\right) = \begin{cases}& \theta-t \hspace{2mm} \text{if} \hspace{2mm} t \geq \theta \\& 0 \hspace{9mm}  \text{else} \hspace{8mm} \end{cases}. \label{hockey2} 
    \end{equation}
Natural data series can in general not be modeled by such a simple behavior as given by these functions. Therefore we 
add some random fluctuations $\xi$ around the mean behavior. These random fluctuations can be due to measurement 
noise as well as to some intrinsic variability, which is not captured by the low dimensional mean dynamics on both sides
of the change point $\theta$. For this fluctuating part of the signal we suppose that its amplitude is 
essentially constant around the change point. The intrinsic variability however may, like  the mean behavior of the system itself,
undergo a sudden change  in its evolution of amplitude. 
Hence we consider stochastic fluctuations $\xi(t)$ whose amplitudes undergo a transition themselves according to
    \begin{equation}
\mbox{STD} (\xi(t)) = \sigma (1 \,+\, s_{1} |t-\theta|_{-} \,+\, s_{2} |t-\theta|_{+})\,.
    \end{equation}
The scale factor $\sigma$ could be the level of the measurement noise or some background level of the intrinsic 
fluctuations, whereas the constants $s_{1,2}$ describe the systematic evolution of the models intrinsic variability 
prior and after the change point measured in units of $\sigma$. 
Although clearly the fluctuating part may contain coherent parts, we assume that throughout this work, 
that the fluctuations are Gaussian random variables, wich at different time points are uncorrelated
    \begin{equation}
\mathbb{E}(\xi(t)\xi(t^\prime)) = 0\,,\quad t\not=t^\prime\,.
    \end{equation}
This clearly is an approximation and its validity can be questioned in concrete applications. However
this assumption  allows us to implement highly efficient algorithms for the estimation of the involved parameters.
From now on we will call this fluctuating part simply ``noise''.
A realization of such a time series is presented in Fig.{\bf\ref{fig:1}}.
Given a data set of $n$ time points $t_i$, $i=1,\dots,n$, the observation vector $\bm{y} =[s(t_i)]^t \in \mathbb{R}^{n}$ can be written
as follows
    \begin{equation}
      \bm{y} = F\bm{\beta} + \bm{\xi}\hspace{2mm}.\label{lm}
    \end{equation}
Here the fixed effect vector $\bm{\beta}= (\beta_{0},\beta_{1},\beta_{2})^{T}\in \mathbb{R}^3$ corresponds to the coefficients of the
linear combination of the Hockey sticks modeling the mean behavior.
The system matrix of the fixed effects, $F \in \mathbb{R}^{n\times 3}$, is then given by the sampling of the Hockey sticks 
$\bm{\zeta}^{\theta}_{\pm}$ defined in Equ.(\ref{hockey1},\ref{hockey2}) at the observation points
    \begin{equation}
    F_{\theta} = \left( \begin{array}{ccc} 1 & \hspace{2.5mm} \left(\zeta^{\theta}_{-}\right)_{1} & \hspace{2.5mm} \left(\zeta^{\theta}_{+}\right)_{1}\vspace{1mm}
\\ \vdots & \hspace{2.5mm} \vdots & \hspace{2.5mm} \vdots\vspace{1mm} \\ 1 & \hspace{2.5mm} 
\left(\zeta^{\theta}_{-}\right)_{n} & \hspace{2.5mm} \left(\zeta^{\theta}_{+}\right)_{n}\vspace{1mm} \end{array}\right),
\quad \left(\zeta^{\theta}_{\pm}\right)_{i} = \left(\zeta^{\theta}_{-}\right)(t_i)\,.
    \end{equation} 
The noise $\bm{\xi}\in \mathbb{R}^n$ is a Gaussian random vector with zero mean and covariance matrix 
$\sigma^2\Omega \in \mathbb{R}^{n\times n}$,
    \begin{equation}
     \bm{\xi} \sim \mathcal{N}\left(0,\sigma^{2} \Omega \right). \label{error}
    \end{equation}
The covariance itself is structured noise, which is parametrized by the two slope parameters $\bm{s}=(s_{1},s_{2})$ 
and the change point $\theta$ itself as
    \begin{equation}
     \left(\Omega_{\theta,s_{1},s_{2}}\right)_{ij} = \left(\left[ 1 + s_{1}\left(\zeta^{\theta}_{-}\right)_{j} + s_{2}\left(\zeta^{\theta}_{+}\right)_{j}\right]^{2}\right)\cdot\delta_{ij}\,. \\ \label{covariancematrix} 
    \end{equation}
In conclusion, the probability density of the observations for fixed parameters (i.e. fixed effects, change point, 
slope parameters) can be written as
    \begin{equation}
     \bm{y} \sim \mathcal{N}\left(F\hat{\bm{\beta}},\sigma^{2} \Omega\right)\,.
    \end{equation}
The Likelihood function of the parameters given the data can then be written as
    \begin{equation}
     \mathcal{L}(\bm{\beta},\sigma,\bm{s},\theta|\bm{y})= \frac{1}{(2\pi\sigma^{2})^{\frac{n}{2}}\sqrt{|\Omega|}}\, 
     e^{-\frac{1}{2\sigma^{2}}(\bm{y} - F\bm{\beta})^T\Omega^{-1} (\bm{y} - F\bm{\beta}) }\,.\label{Likelihood}
    \end{equation}
Note that the functional dependency of $\bm{\beta}$ is a Gaussian density. Clearly in the exponential $\bm{\beta}$ 
is of a quadratic form and since $\Xi=F^T\Omega^{-1}F$ is positive definite we may write 
    \begin{equation}
  \frac{1}{(2\pi\sigma^{2})^{\frac{n}{2}}\sqrt{|\Omega|}}\,  e^{-\frac{\mathcal{R}^2}{2\sigma^2}} \,
     e^{-\frac{1}{2\sigma^{2}}(\bm{\beta} - \bm{\beta^\ast})^T \Xi (\bm{\beta} - \bm{\beta^\ast} ) }\label{Likelihood_clear}
    \end{equation}
where the mode of the Gaussian in $\bm{\beta}$ is the best linear unbiased predictor of the fixed effects (BLUP) \cite{Robinson_1991}
\begin{eqnarray}\label{eq:blup}
    \bm{\beta}^\ast &=& \underset{\bm{\beta} \in \mathbb{R}^3}{\mbox{argmin}} (\bm{y} - F\bm{\beta})^T\Omega^{-1} (\bm{y} - F\bm{\beta}) \nonumber
                         \\ &=& (F^T\Omega^{-1}F)^{-1} F^T \Omega^{-1} \bm{y}
\end{eqnarray}
and the residuum $\mathcal{R}$ measured in the Mahalanobis distance \cite{Mahalanobis_1936}, induced by the covariance matrix $\Omega$, is 
\begin{eqnarray}
    \mathcal{R}^2 &=& \min_{\bm{\beta} \in \mathbb{R}^3} (\bm{y} - F\bm{\beta})^T\Omega^{-1} (\bm{y} - F\bm{\beta})\nonumber\\
        &=& (\bm{y} - F\bm{\beta^\ast})^T\Omega^{-1} (\bm{y} - F\bm{\beta^\ast})\,.
\end{eqnarray}
In addition, the profiled Likelihood function $\mathcal{L}(\bm{\beta}^\ast,\sigma,\bm{s},\theta|\bm{y})$ enables us to derive 
the profiled Likelihood estimator of the scale parameter $\sigma$ 
\begin{equation}\label{eq:sigest}
    \hat\sigma^{2} = \frac{\mathcal{R}^2}{n+1}\, ,
\end{equation}
which is auxiliary for the computation of the maximum of the Likelihood function.


\subsection{Bayesian inversion}
\label{inversion}

    \begin{figure*}[t]
      \begin{tabular}{lcr}
       \hspace{-6mm} a)\hspace{-1.3mm} \includegraphics[width=0.34\textwidth]{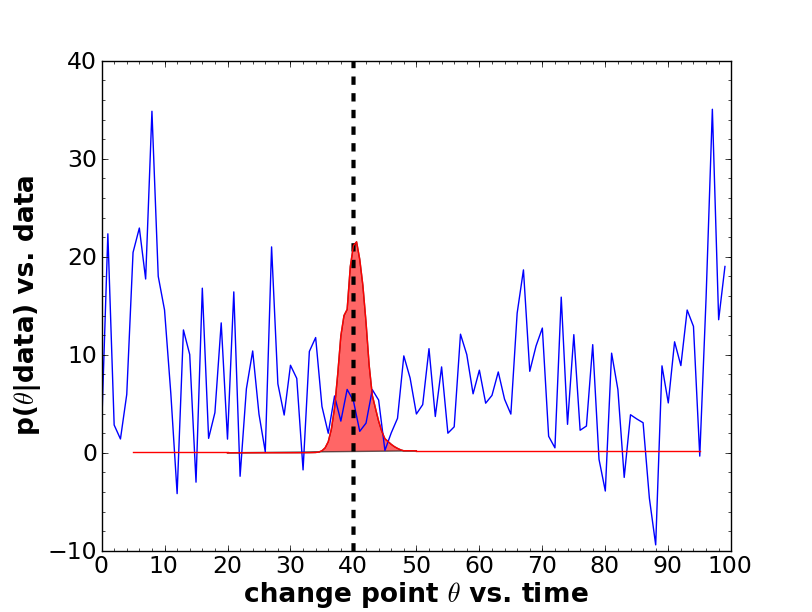} & \hspace{-3mm} b) \hspace{-2.3mm} \includegraphics[width=0.32\textwidth]{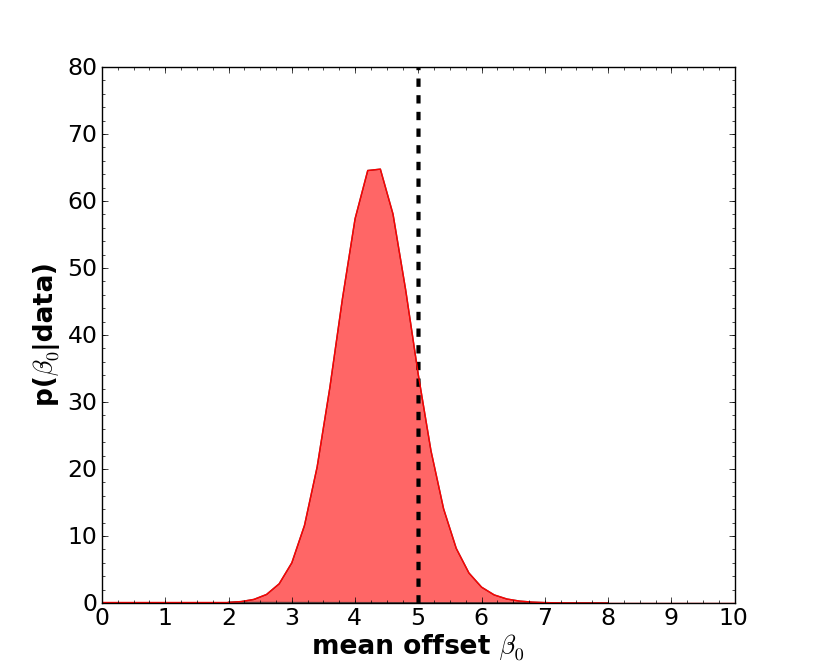}
       & \hspace{-3mm} c) \hspace{-2.5mm} \includegraphics[width=0.335\textwidth]{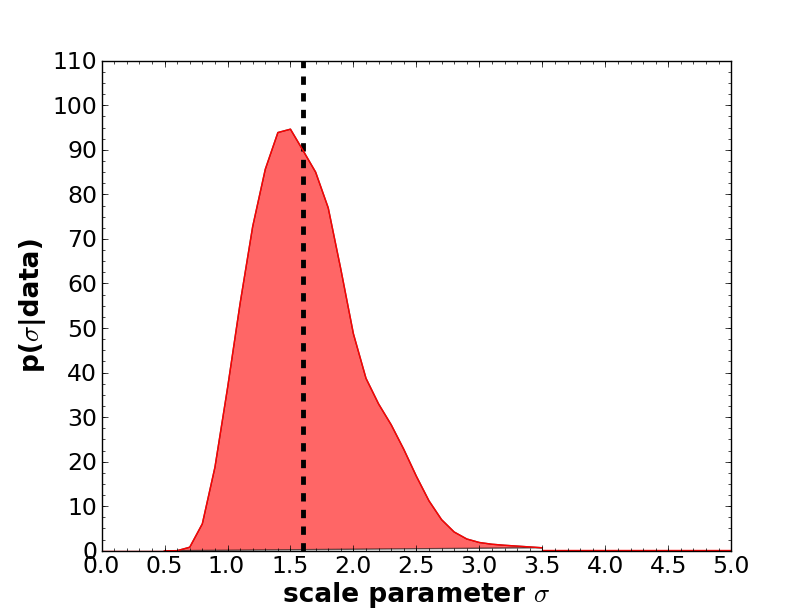}
      \end{tabular}
	\caption{Normalized marginal posterior densities for the time series in Fig.~{\bf\ref{fig:1}}. 
                 The maxima indicate the most probable estimates of the
		 a) change point $\hat\theta=40.5$, b) fixed effect offset $\hat\beta_{0}=4.40$ and  
                 c) scale parameter $\hat\sigma=1.50$. The dashed lines represent the true parameter values of the underlying model.}
	\label{fig:2}       
    \end{figure*}
 
In the light of the Bayesian theorem, we can compute the posterior distribution $p(\bm{\beta}, \sigma, \theta, \bm{s}|\bm{y})$ 
of the modeling parameters given the data $\bm{y}$  from the Likelihood function  
Eq.~(\ref{Likelihood}) by specifying the prior distribution of the parameters $p(\bm{\beta}, \sigma, \theta, \bm{s})$, 
which encodes our belief about the parameters prior to any observation. 
Since we assume a priori no correlations between the parameters, the joint prior distribution can be factorized into 
the independent parts
    \begin{equation}
      p(\bm{\beta}, \sigma, \theta, \bm{s}) =  p(\theta) \cdot p(\bm{s}) \cdot p(\sigma) \cdot p(\bm{\beta})\,.
    \end{equation}
In general, we do not have any a priori knowledge about these  hyperparameters  and thus we shall use flat and
uninformative priors \cite{Raftery_1986,Wahba_1978}
    \begin{equation}
     p(\theta) \sim 1\,,\hspace{2mm} p(\bm{s}) \sim 1 \,,\hspace{2mm} p(\bm\beta) \sim 1\,,
    \end{equation}
For the scale parameter $\sigma$ we assume a Jeffrey's prior \cite{Jeffreys_1946}
    \begin{equation}
     p(\sigma) \sim \frac{1}{\sigma}\,. \label{jeffrey}
    \end{equation}
These statisitical assumptions enable us to compute the posterior density of the system's parameters given the data 
$\bm{y}$ as
    \begin{equation}
      p(\bm{\beta}, \sigma, \theta, \bm{s}| \bm{y}) = C \cdot \mathcal{L}(\bm{\beta}, \sigma, \theta, \bm{s}| \bm{y}) \cdot \frac{1}{\sigma}\,. \label{postplain}
    \end{equation}
The normalization constant $C$ ensures that the right hand side actually defines a normalized probability density.
From this expression, various marginal posterior distributions may be obtained by integrating over the parameters 
that shall not be considered.
We are mostly interested in the posterior distribution of the possible change point locations $\theta$. 
To produce the posterior distribution of this quantity, we have to marginalize out all other variables. 
It turns out that all but the integral over the noise slopes $\bm{s}$ may be carried out explicitely. 
Thanks to the Gaussian nature of the $\bm{\beta}$ dependency we obtain
   \begin{equation}
     p(\sigma,\theta,\bm{s} | \bm{y}) \sim \frac{\sigma^{1-n}}{\sqrt{|\Omega||F^T\Omega^{-1}F|}} e^{-\frac{1}{2\sigma^{2}}\mathcal{R}^{2}}\label{sigma} \, ,
   \end{equation} 
and
   \begin{equation}
     p(\bm{\beta},\theta,\bm{s} | \bm{y}) \sim \frac{\left[(\bm{y} - F\bm{\beta})^T\Omega^{-1} (\bm{y} - F\bm{\beta})\right]^{-\frac{n}{2}} }{\sqrt{|\Omega|}} \label{beta}\, .
   \end{equation} 
Further marginalization may be performed to yield
    \begin{eqnarray}
      p(\theta, \bm{s} | \bm{y}) &=&  \int d\sigma d\bm{\beta}\ p(\bm{\beta},\sigma,\theta, \bm{s}| \bm{y})\\
     &=& C^\prime \cdot \frac{\mathcal{R}^{-(n-2)}}{\sqrt{|\Omega||F^T\Omega^{-1}F|}}\,.\label{pthetaslopes}
    \end{eqnarray}
Again $C^\prime$ is a constant, that ensures the normalization of the right hand side to a probability density.
Finally the posterior marginal distribution of $\theta$ can be computed by numeric evaluation of the following integral
    \begin{equation}
p(\theta | \bm{y}) = \int d\bm{s}\ p(\theta, \bm{s} | \bm{y})\,.\label{ptheta}
    \end{equation}
In the same way the numeric $\theta$ integral may be performed to elaborate the posterior information about the 
involved slope parameters $\bm{s}$ of the heteroscedastic behavior around the change point
    \begin{equation}
p(\bm{s} | \bm{y}) = \int d\theta\ p(\theta, \bm{s} | \bm{y})\,.\label{pslope}
    \end{equation}


\section{Validation the method}
\label{performance}

In order to validate the method's performance in an idealized setting we use synthetic time series to discuss 
its ability to estimate the model's parameters and to elaborate the sensitivity of the estimates to data loss. 
We generate the time series via the LMM Equ.(\ref{lm}) and infer on the change point by computing the global marginal 
posterior density Equ.(\ref{ptheta}), i.e. over the interval of all possible change point values $\theta$. 
The location of the maximum of the marginal posterior density $\left[p(\theta|\bm{y})\right]_{\mbox{max}}$ can be used 
as an estimator for the most probable location of a singularity $\hat\theta$.  In case, the data contains more than one 
change point, the posterior distribution will exhibit multiple local maxima. This could therefore be used as an indicator 
for the existence of secondary change points in the time series. Although a more reasonable way would be to consider 
models with multiple change points this approach becomes quickly uncomputable due to exploding dimensionality. 
Thus we propose a local kernel based method to be able to apply our single change point model locally to multi change 
point data series.

\subsection{Estimation of a single change point}
\label{single}

    \begin{figure*}
      \begin{tabular}{lr}
       a) \includegraphics[width=0.4\textwidth]{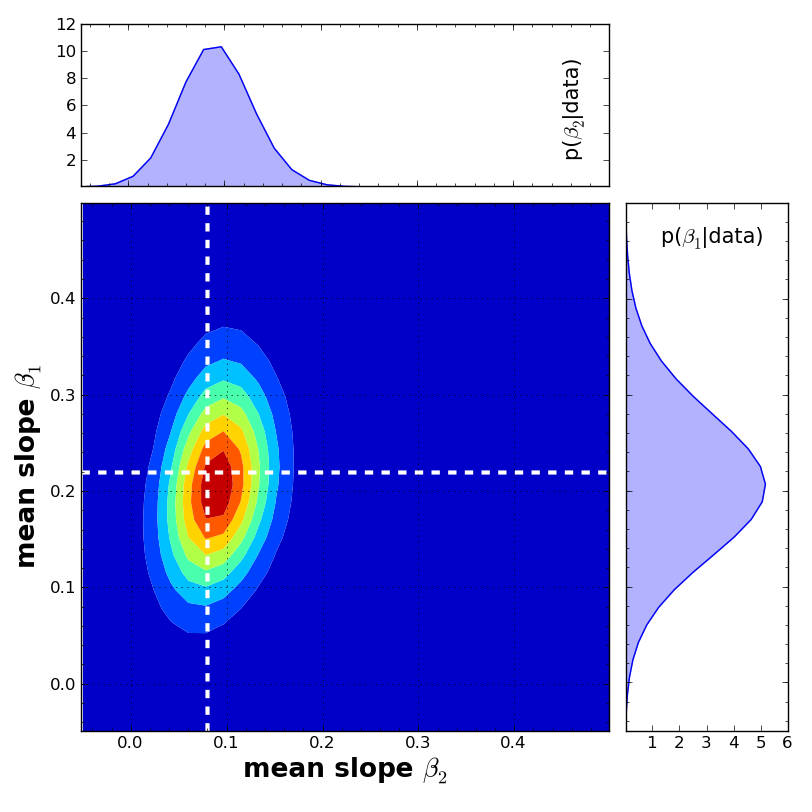} & \hspace{1cm} b)  \includegraphics[width=0.4\textwidth]{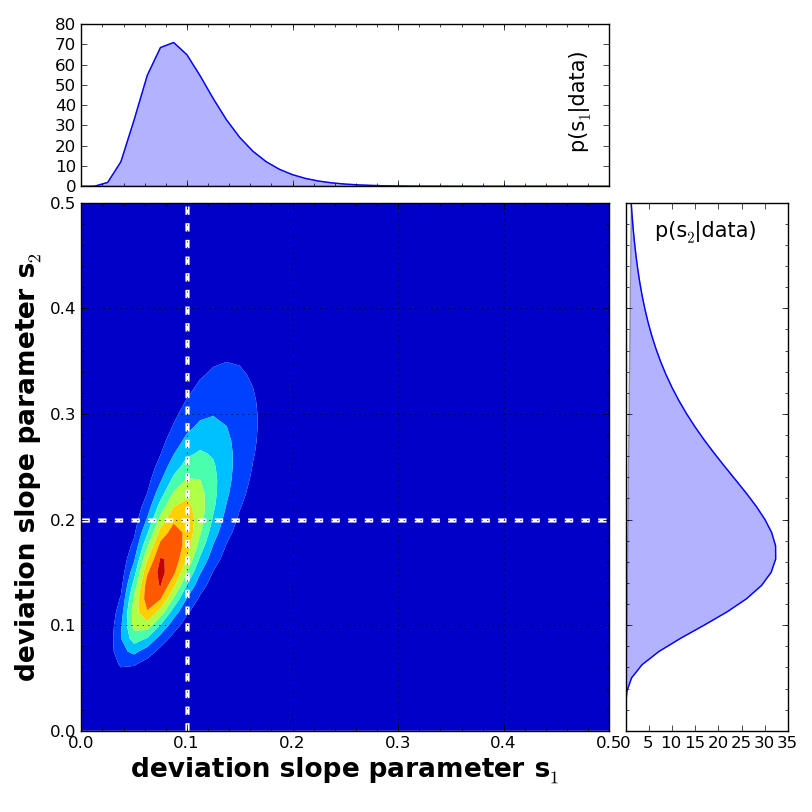}
      \end{tabular}
	\caption{Normalized two dimensional marginal posterior densities for the time series in Fig.{\bf\ref{fig:1}}. 
                 The maxima indicate the most probable estimates of the 
                 a) fixed effect slopes $(\hat\beta_{1},\hat\beta_{2})=(\,0.206\,,\,0.096\,)$ and b) 
                 deviation slope parameters $(\hat s_{1},\hat s_{2})=(\,0.087\,,\,0.167\,)$. 
                 Alongside the contour plots are presented the one dimensional projections of the posterior densities.
                 The dashed lines represent the true values of the underlying model.}
	\label{fig:3}       
    \end{figure*}
To validate our technique, we apply it to the generated time series of Fig.{\bf\ref{fig:1}} containing a single 
change point at $\theta=40$. We compute all relevant two and one dimensional marginal distributions of the 
model's parameters using the formulas of the previous section.
The marginal distributions provide Bayesian estimates for the change point $\theta$, 
mean behavior $\bm{\beta}$, scale parameter $\sigma$ and heteroscedastic behavior $\bm{s}$ of the data 
as the maxima of the one and two dimensional marginal distributions shown in Fig.{\bf\ref{fig:2}}, {\bf\ref{fig:3}}.\\ 
First note that due to the random nature of the observations, the posterior density too depends randomly on the actual series 
of observations. It is therefore not surprizing, that the locations of the maxima of the posterior does 
not exactly agree with the true parameter values. However, they are within a certain quantile of the posterior distribution. 
We automatically obtain confidence intervals or regions by considering those level intervals or contour-lines, 
that enclose a fixed percentage of the total probability. This yields a natural way of uncertainty quantification.\\
The estimated change point $\hat\theta=40.5$ differs only little from the real value $\theta=40.0$ within a 
relatively narrow and symmetric confidence interval $[\,35.7\,,\,45.9\,]$ (Fig.{\bf\ref{fig:2}}a). Consequently we achieve 
to restrict the location of a probable singularity to a range $<9\%$ of the time grid. 
The estimates of the mean behavior are obtained from Fig.{\bf\ref{fig:2}}b, {\bf\ref{fig:3}}a as
$\hat{\bm{\beta}}=(\,4.40\,,\,0.206\,,\,0.096\,)$. The Bayesian estimates reproduce the real 
underlying mean model $\bm{\beta}=(\,5.0\,,\,0.22\,,\,0.08\,)$ convincingly. The two dimensional 
contour plot of the marginal density $p(\beta_{1},\beta_{2}|\bm{y})$ of the mean slopes 
indicate an approximate symmetric confidence area of the most probable slope combinations $(\beta_{1},\beta_{2})$ 
(red area in Fig.{\bf\ref{fig:3}}a). 
The one dimensional projection $p(\beta_{1}|\bm{y})$ reveals a broader confidence interval for the estimation of 
$\hat{\beta_{1}}$ compared to $\hat{\beta_{2}}$. 
The scale parameter can be estimated as $\hat{\sigma} = 1.50$ from Fig.{\bf\ref{fig:2}}c within the confidence interval 
$[\,0.806\,,\,2.84\,]$ unidirectional wider to growing $\sigma$-values and differs little from the true value 
$\sigma = 1.60$. 
The two dimensional contour plot of the marginal density of the deviation slope parameters $p(s_{1},s_{2}|\bm{y})$ 
indicate a slight asymmetric confidence area of the most probable slope combinations $(s_{1},s_{2})$ 
(red area in Fig.{\bf\ref{fig:3}}b). The one dimensional projections $p(s_{1}|\bm{y})$ and $p(s_{2}|\bm{y})$ display  
unidirectional wider confidence bounds for the estimates $\hat{s_{1}}=0.087$ to bigger and $\hat{s_{2}}=0.167$ to smaller 
parameter values.
\begin{table}[h]
\caption{Estimated model of the synthetic signal of Fig.~{\bf\ref{fig:1}}}
\label{tab:1} 
\begin{tabular}{cll}
\hline
\hline\noalign{\smallskip}
parameter \hspace{2mm} & estimate \hspace{2mm} & confidence $\geq95\%$  \\
\noalign{\smallskip}\hline\noalign{\smallskip}
$\hat\theta$ & 40.5 & [$\,$35.7$\,$,$\,$45.9$\,$] \\
$\hat\beta_{0}$ & 4.40 & [$\,$3.00$\,$,$\,$5.75$\,$] \\
$\hat\beta_{1}$ & 0.206 & [$\,$0.035$\,$,$\,$0.390$\,$] \\
$\hat\beta_{2}$ & 0.096 & [$\,$-0.015$\,$,$\,$0.189$\,$] \\
$\hat\sigma$ & 1.50 & [$\,$0.806$\,$,$\,$2.84$\,$] \\
$\hat s_{1}$ & 0.087 & [$\,$0.027$\,$,$\,$0.220$\,$] \\
$\hat s_{2}$ & 0.167 & [$\,$0.050$\,$,$\,$0.380$\,$] \\
\noalign{\smallskip}\hline
\hline
\end{tabular}
\end{table}\\
Thus for our realization, the marginal distributions of the heteroscedastic behavior $(\sigma,s_{1},s_{2})$ indicate a broad range of probable 
parameter combinations compared to the mean behavior $\bm\beta$ or the change point $\theta$.
In Tab.{\bf\ref{tab:1}} we summarize our point estimators and $95\%$ confidence intervals for them based on our 
analysis. 

\subsubsection{Sensitivity to data loss}
\label{sensitivity}

\begin{figure}
        \includegraphics[width=0.5\textwidth]{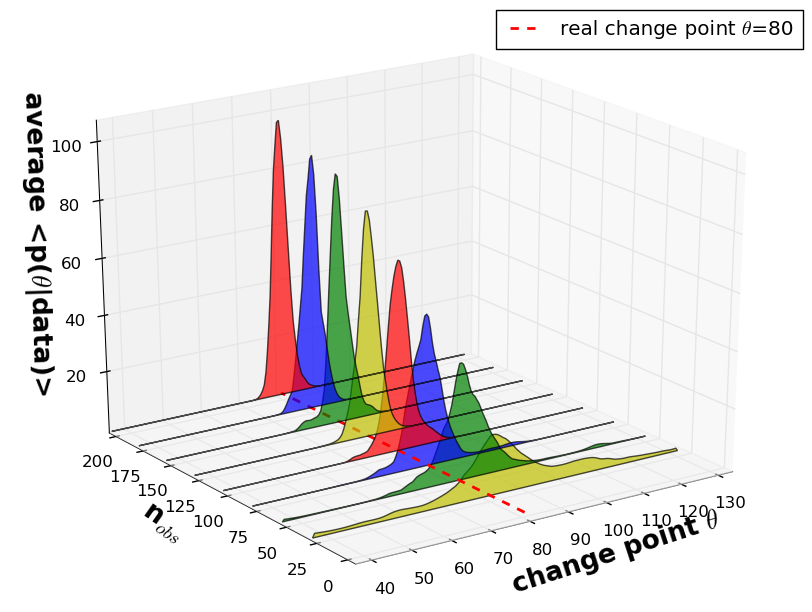}
	\caption{The global maxima of the averaged posterior densities $\left<p(\theta| \bm{y})\right>_{n_{obs}}$ 
                 converge for increasing number of data points $n_{obs}$ towards a delta distribution located at the true 
                 change point value $\theta = 80$.}
	\label{fig:4} 
    \end{figure}

\begin{figure}
        \includegraphics[width=0.53\textwidth,height=0.35\textwidth]{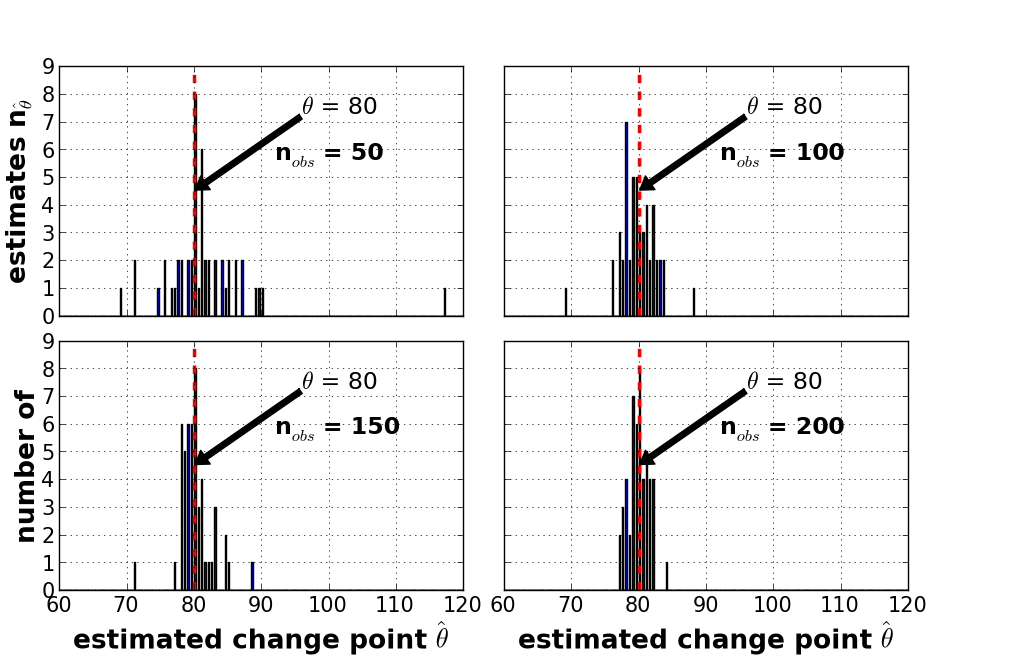}
	\caption{Histograms of the global change point estimators $\hat\theta_{obs}^{i}$ for $i=1,\dots,50$ realizations 
                 and with respect to 
                 $n_{obs}=50,100,150,200$ data points from the setting of Fig.{\bf\ref{fig:4}}. 
                 Even for $n_{obs}=100$ nearly all global estimates $\hat\theta^{i}_{100}$ lie in the 
                 interval $[76.0\,,\,83.5]$, respectively $<4\%$ of the full time grid.}
	\label{fig:5} 
    \end{figure}

In real data, analysts have to deal with sparse and irregularily sampled data. Our technique does not require an uniform 
sampling grid of data points since from the beginning, it employs only the available data. 
As a validation for the sensitivity of our method to data loss, we randomly ignore stepwise $0\%$ up to $87,5\%$ of 
the time series modeled by a sequence of $n_{obs}=200$ observations. The artifical time series undergo a change point 
$\theta = 80$ and are further parametrized by the mean
$F_{\theta}\bm{\beta}= 12 + 0.24\cdot\bm{\zeta}^{\theta}_{-} + 0.02\cdot\bm{\zeta}^{\theta}_{+}$ 
and the deviation behavior $\sigma^{2} \Omega_{\theta, \bm{s}}= \left[1.2(1 + 0.18\cdot\bm{\zeta}^{\theta}_{-} + 0.04\cdot\bm{\zeta}^{\theta}_{+})\right]^{2}$. 
Leaving out randomly a defined percentage of the observations produces time series with random gaps and irregular
sampling steps. For each of these random realizations consisting of $n_{obs}$ data points we compute the posterior 
densities $p_{n_{obs}}^{i}(\theta | \bm{y})$ for $i=1,\dots,50$ realizations.\\
The obtained averaged posterior densities $\left<p(\theta | \bm{y})\right>_{n_{obs}}$ in the plane of the sample size $n_{obs}$ 
are shown in Fig.{\bf\ref{fig:4}}, indicating with their maxima the averaged most probable change points $\hat{\left<\theta\right>}_{n_{obs}}$.
Apparently the mean of the posterior densities differs from the true value, however still within the
width of the distribution. The latter depends invers proportionaly on the square root of the sample size
\begin{equation}
\mbox{width}\left[\left<p(\theta | \bm{y})\right>_{n_{obs}}\right] \propto \frac{1}{\sqrt{n_{obs}}} \hspace{3mm} .
\end{equation}
At large numbers of sampling points $n_{obs}$ the posterior converges towards a delta distribution located at the true 
parameter value $\theta=80$. In any case, even for small data sets, as small as $n_{obs}=25$, the non-flatness of the 
posterior clearly hints towards the existence of a change point in the time series.
The investigation of the averaged marginal posterior densities in the plane of the remaining parameters reveals 
a broadening of the posterior distributions for $n_{obs}<200$, as naturally expected due to information loss in the 
sub time series considered in the inference process.\\
Additionally we point out the efficiency of our method to infer on the explicit location of a singularity $\hat\theta_{n_{obs}}^{i}$ 
for every single time series of the previous setting. In Fig.{\bf\ref{fig:5}} are presented the histograms of the global 
point estimators $\hat\theta_{n_{obs}}^{i}$ for every single realization $i=1,\dots,50$. 
We observe that the particular global estimators $\hat\theta_{obs}^{i}$ are relatively robust to data loss and enable us 
to infer convincingly on the location of the singularity. Even considering only $50\%$ of the full time series, i.e.  
$n_{obs}=100$, produces global estimates that lie in the narrow interval $[76.0\,,\,83.5]$, representing $<4\%$ of the 
full time grid.\\   
However, for such a data-poor situation, local additional, less dominant maxima are likely to appear due to random 
fluctuations in the posterior, and more sofisticated techniqes are needed to assess the existence of single or multiple
change points. 
One approach to clearify multimodial posterior densities is the computation of local posterior 
densities within a sliding window as presented in the following.

\subsubsection{Local posterior density}
\label{localposteriori}

\begin{figure}
        \includegraphics[width=0.48 \textwidth]{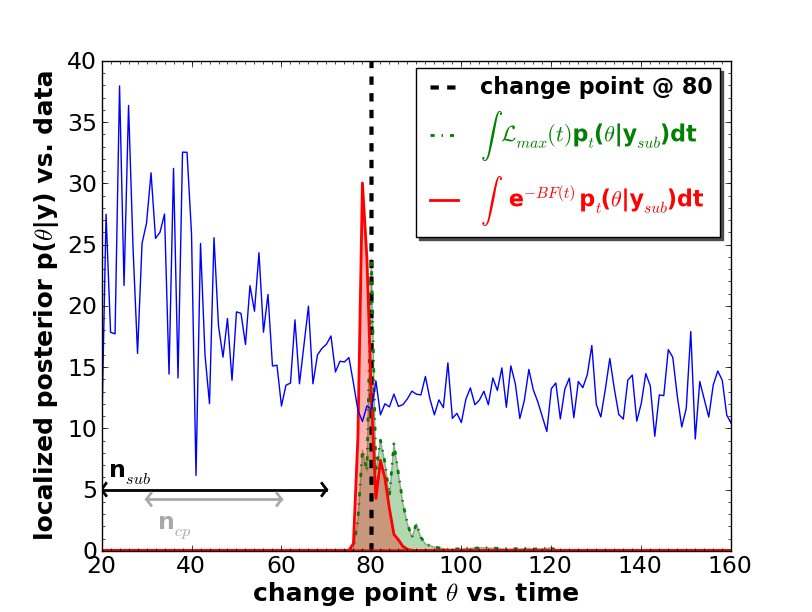}
	\caption{Normalized sum of local posterior densities weigted by the local Likelihood maxima (dashed) and with respect to the Bayes factor 
                 (solid), computed for sub time series of $n_{sub}=50$ data points 
                 and a sampling grid of $n_{cp}=30$ change points. 
                 The data is one realization of the time series 
                 defined in Sect.\ref{sensitivity} for $n_{obs}=200$ data points.}
	\label{fig:6} 
    \end{figure}

\begin{figure}
        \includegraphics[width=0.48\textwidth]{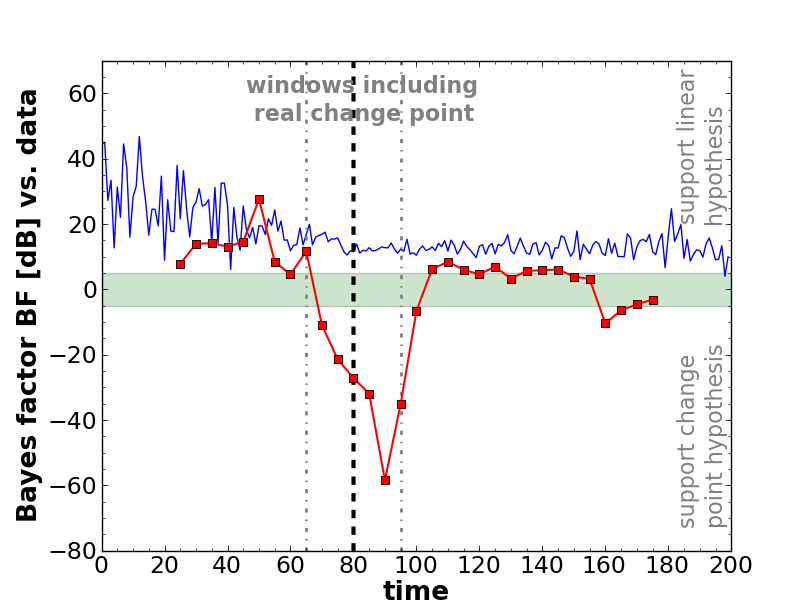}
	\caption{Local Bayes factor (squares) obtained for the time series in Fig.{\bf\ref{fig:6}}. 
                 The shaded area encloses values whose support for none of the models is 
                 substantial (based on \cite{Raftery_1995}). Values underneath this area strongly support a change point against a linear model,
                 and vice versa for values above.}
	\label{fig:7} 
    \end{figure}

\begin{figure}[b]
	\includegraphics[width=0.48\textwidth]{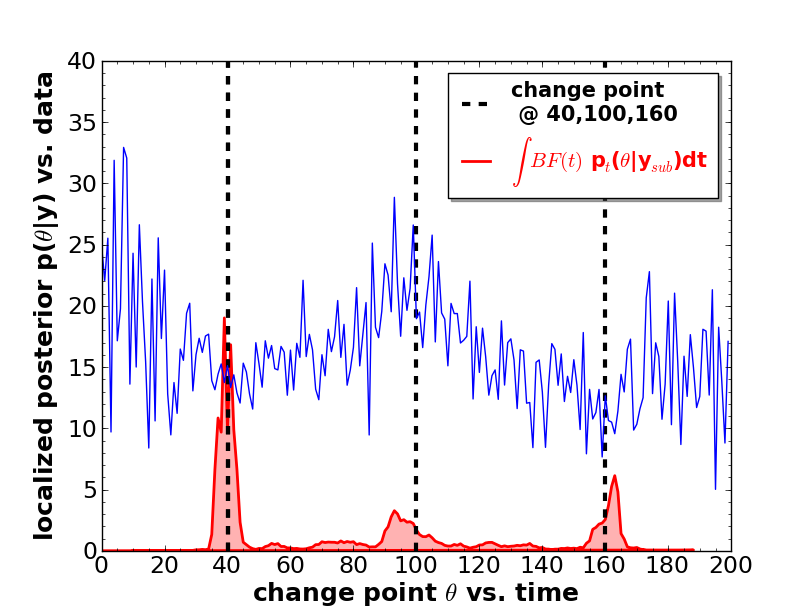}
	\caption{Normalized sum of local posterior densities weighted by the Bayes factor, computed for sub time 
                 series of $n_{sub}=50$ data points 
                 and a sampling grid of $n_{cp}=30$ change points. The 
                 parametrization of the mean is defined as
                 $F\bm{\beta}= 14 + 0.2\cdot\bm{\zeta}^{40}_{-} + 0.1\cdot\bm{\zeta}^{40}_{+} - 0.25\cdot\bm{\zeta}^{100}_{+} + 0.3\cdot\bm{\zeta}^{160}_{+}$ 
                 and the deviation is modeled as
                 $\sigma^{2} \Omega= \left[1.6(1 + 0.2\cdot\bm{\zeta}^{40}_{-} + 0.03\cdot\bm{\zeta}^{40}_{+} - 0.05\cdot\bm{\zeta}^{100}_{+} + 0.1\cdot\bm{\zeta}^{160}_{+})\right]^{2}$.}
	\label{fig:8}       
    \end{figure}

Long data sets are likely to contain more than one change point. So using our model globally may not be justified. However, locally our model
assumption may still be valid. For this reason, we propose the following kernel based local posterior method. In addition this method allows us to treat very long
data sets numerically more efficient since the computation scales with the the third power of the employed data points.
Around each time point $t$ we choose a data window $I_t=[t-\frac{T}{2}, t+\frac{T}{2}]$ of length $T$. 
Inside this window, we take as prior distribution for the change point location $p(\theta)$ a flat prior inside 
some subinterval of length a:
\begin{equation}
p(\theta)=\begin{cases}\frac{1}{a} &\hspace{1mm}\mbox{for} \hspace{3mm} t-\frac{a}{2}\leq \theta \leq t+\frac{a}{2}\\ 
                                 0 &\hspace{1mm}\mbox{else} \end{cases}\,,\quad 0<a<T\,.
\end{equation}
We then compute the local posterior $p_t(\theta|\bm{y}_{|I_t})$ around $t$  based on the subseries in the data window 
$\bm{y}_{|I_t}$. This yields a posterior distribution of a possible change point within each window under the 
assumption that there is actually a singularity within the window. In order to compare different
window locations, we need to quantify the credibility that there is a change point. Therefore we compute the maximum 
of the Likelihood within each window
\begin{equation}
f(t)= \underset{\theta\in[t-\frac{1}{a},t+\frac{1}{a}],s_1,s_2 \in \mathbb{R}}{\mbox{max}}\mathcal{L}( \bm{\beta}^\ast, \hat\sigma; \bm{y}_{|I_t}),
\end{equation}
where $\hat\sigma$ and  $\bm{\beta}^\ast$ are the estimators given by Eq.(\ref{eq:sigest}) and (\ref{eq:blup}). 
The global distribution of change points $\theta$ given the full time series is then obtained as a weighted superposition 
in form of
\begin{equation}
p(\theta|\bm{y}) = C\cdot\int f(t) \, p_t(\theta| \bm{y}_{|I_t}) dt\,,\label{weighting}
\end{equation}
whereas the constant $C$ ensures the normalization to a probability density.
In subdata sets with no change point, the credibility of the model fit is very low, 
in conclusion the Likelihood maxima is of very small value and local estimates are judged as negligible. 
By construction the method works for multiple change points as soon as they are separated by at least one data window.
We demonstrate this by applying our algorithm first on a synthetic single change point time series. 
In Fig.{\bf\ref{fig:6}} is shown the sum of the local posterior densities weighted by the maxima of the local 
Likelihood (dashed curve). The time series is one realization of the model in the previous Sect.\ref{sensitivity}
for a sequence of $n_{obs}=200$ data points. Supplementary the applied window size $n_{obs} = 50$ and the 
sampling grid of the change points $n_{cp} = 30$ are presented for comparison. 
The sum of local posterior densities indicates the best model fit for windows covering the real change point $\theta=80$ 
but is non-zero even between $[100, 121]$ suggesting that a change point model might be suitable for these singularity values as well.\\  
A second quantity that may be used to produce relative credibility weights for the windows 
is given by the Bayes factor \cite{Raftery_1995}. 
Besides the goodness of fit, the complexity of the assumed model has to be taken into account to assess the most 
capable model describing the data and thus performing the estimation. Thus we test the hypothesis of no change point, 
respectively a linear model $\mathcal{M}_{lin}$, against a change point model $\mathcal{M}_{cp}$ in form of the Bayes factor
    \begin{equation}
     BF(t) = \frac{p\left(\mathcal{M}_{lin}| \bm{y}_{|I_t}\right)}{p\left(\mathcal{M}_{cp}|\bm{y}_{I_t}\right)}\,.
    \end{equation}  
The dependency of the Bayes factor on a logarithmic scale is shown in Fig.{\bf\ref{fig:7}} for the artifical time series 
of Fig.{\bf\ref{fig:6}}. 
The Bayes factor in this test case favors the change point over the linear model for all local  windows, for which 
the true change point is in the support of the inner  prior distribution of $\theta$. This local Bayes factor itself 
can be used as a diagnostic tool like the Likelihood weighted posterior,
but we may also combine the techniques by using the $BF$ as a window weighting function by setting $f(t)=e^{-BF(t)}$ 
in Eq.(\ref{weighting}). 
In this form Eq.(\ref{weighting}) corresponds therefore essentially to the total probability decomposition of the change point (cp)
    \begin{equation}
        \sum_{\small\mbox{windows}} p(\theta | \mbox{cp in window})\, p(\mbox{cp exists in window})\,.
    \end{equation}  
For comparison of both kernel approaches we present in Fig.{\bf\ref{fig:6}} additionally the sum of local posterior densities 
weighted by $e^{-BF(t)}$ (solid curve). The distribution weighted with respect to the Bayes factor are non-zero 
in the range between $[78,89]$ whereas the one weighted by the maxima of the Likelihood is non-zero in $[78,121]$. The long tail 
of the latter hints to less probable change point locations which are automatically rejected in the Bayes factor 
weighting.
Furthermore we exemplify the algorithm on a synthetic multi change point time series shown in Fig.{\bf\ref{fig:8}}.
For clarity of presentation we plot the sum of posterior distributions weighted with the plain Bayes factor $BF$.  
We are able to infer on the true change point values $(\theta_{1},\theta_{2},\theta_{3})=(40,100,160)$ via the 
estimators $(\hat\theta_{1},\hat\theta_{2},\hat\theta_{3})=(38.9,93.0,162.9)$ within their intervals 
$([33.9,47.8],[87.5,109.1],[158.9,167.0])$ of about $90\%$ confidence. We obtain these intervals from a more detailed 
analysis of the partial sums of local posterior densities weighted by the factor $e^{-BF}$ covering the estimated 
singularity locations.\\
The main advantage of this localization approach even in a single change point context is however the enormous 
speedup of the computations. For instance for a time series of $n_{obs}=2000$ data points we pass from a global computation of the 
marginalized posterior density in $3h\,41min\,40s$ to 
a local one divided into 40 overlapping subdata sets of $n_{sub}=100$ in $7min\,44s$, respectively a speed up of about $95\%$.  
This is achieved using Python 2.6.5 on a Supermicro Intel(R) Core(TM)i7 CPU 920 @ 2.68GHz with 12GB RAM. 
In the context of complex multiple change point scenarios, as real time series mostly are, the localization approach 
of the posterior density $p(\theta|\bm{y})$ combined with the Bayes factor realizes a powerfull tool to scan the data seperately 
for single change points, as demonstrated in the following Sect.\ref{Nile}.


\subsection{Annual Nile flow from 1871 to 1970}
\label{Nile}

\begin{figure}
	\includegraphics[width=0.5\textwidth]{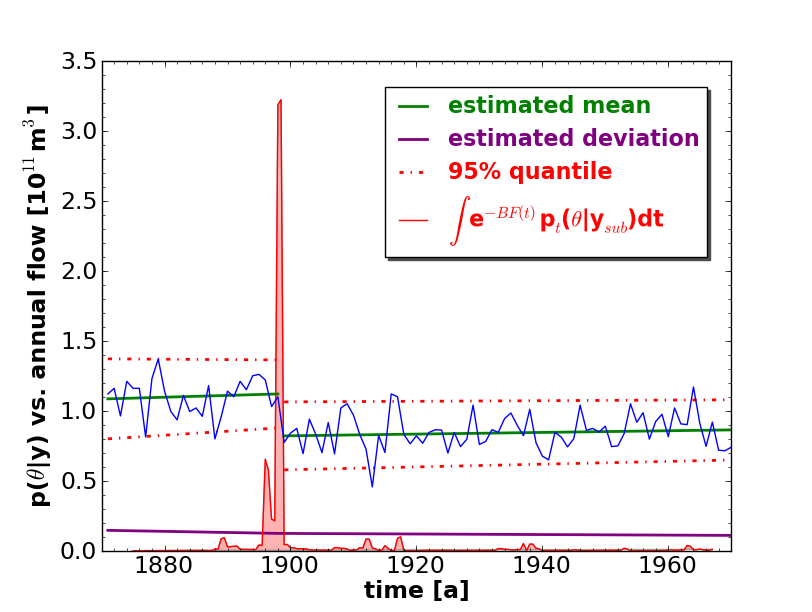}
	\caption{Annual Nile flow containing a known change point at $\theta = 1899$. 
                 The sum of localized posterior densities weighted with respect to the Bayes factor $BF$ 
                 indicates a change point at $\hat\theta = 1898$ within its confidence 
                 interval $[1896,1900]$ of about $90\%$. The estimated underlying model
                 reveals the most dominant transition in the behavior of the mean.}
	\label{fig:9}       
    \end{figure}

We demonstrate our technique by applying it on a time series including a known significant change point. 
For this purpose we analyze the annual Nile River flow measured at Aswan from 1871 to 1970 \cite{Cobb_1978}. 
Several investigation methods have verified a shift in the flow levels starting from the year 1899 \cite{Cobb_1978, 
Jong_1998, Downey_2008}. Historical records provide the fact, that this shift is attributed partly to 
weather changes and partly to the start of construction work for a new dam at Aswan.
Since we expect a natural behavior of the underlying mean we generalize our previous model to undergo besides 
trend changes as well a sharp shift in the mean offset at the singularity $\theta$.
Therefore we modify the system matrix according to  
    \begin{equation}
    F_{\theta} = \left( \begin{array}{cccc} \left(\varphi^{\theta}_{-}\right)_{1} & \hspace{2.5mm} \left(\zeta^{\theta}_{-}\right)_{1} & \hspace{2.5mm} \left(\zeta^{\theta}_{+}\right)_{1} & \hspace{2.5mm} \left(\varphi^{\theta}_{+}\right)_{1} \vspace{1mm}
\\ \vdots & \hspace{2.5mm} \vdots & \hspace{2.5mm} \vdots & \hspace{2.5mm} \vdots\vspace{1mm} \\ \left(\varphi^{\theta}_{-}\right)_{n} & \hspace{2.5mm} \left(\zeta^{\theta}_{-}\right)_{n} & \hspace{2.5mm} \left(\zeta^{\theta}_{+}\right)_{n} & \hspace{2.5mm} \left(\varphi^{\theta}_{+}\right)_{n}\vspace{1mm} \end{array}\right)\hspace{2mm} ,
    \end{equation}
whereas we define another type of Hockey sticks $\bm{\varphi}^{\theta}_{-}$ and $\bm{\varphi}^{\theta}_{+}$ 
referring to Eq.(\ref{hockey1}) and (\ref{hockey2}) not as linear but as constant. The general formulas of the Bayesian inference 
remain the same, with these new functions. 
First of all we compute the global posterior density $p(\theta, \bm{s} | \bm{y})$ as presented in Eq.(\ref{pthetaslopes}). 
By initially guessing a reasonable sampling grid for the change point $\theta$ and the slope parameters $\bm{s}$
from the data, we clearly obtain significant maxima in the posterior projections $p(\theta|\bm{y})$ and $p(\bm{s} | \bm{y})$. 
Therefore we adjust the sampling grid to obtain finer posterior structures around the obvious maxima. We estimate the 
change point as $\hat\theta = 1898$ within a confidence interval $[1895,1901]$ of over $95\%$. 
The slope parameters of the deviation are estimated as $(\hat{s}_{1},\hat{s}_{2}) = (0.0065, -0.0015)$ 
within the $90\%$ confidence intervals $\hat{s}_{1}$ in $[-0.0190,0.0450]$ and $\hat{s}_{2}$ in $[-0.0065,0.0855]$.\\
Prior the estimators $\hat\theta$ and $\hat{\bm{s}}$ we compute the posterior 
projections $p(\bm{\beta}, \theta, \bm{s} | \bm{y})$ and $p(\sigma, \theta, \bm{s} | \bm{y})$ formulated in Eq.(\ref{beta}) and (\ref{sigma}). By minimizing the 
sampling grid of $\theta$ and $\bm{s}$ to its confidence intervals we are able to speed up the compuation and to estimate 
the remaining parameters $\bm{\beta}$ and $\sigma$. Finally we reveal from the global posterior distribution 
the most probable model plotted in Fig.{\bf\ref{fig:9}} and listed in Tab.{\bf\ref{tab:2}}. \\
Additionally we investigate the time series for local singularities by computing the sum of local posterior densities 
weighted by the Bayes factor as $e^{-BF}$ (displayed in Fig.{\bf\ref{fig:9}}) for the window sizes $n_{sub}=50a$ of 
considered subseries. The change point sampling grid contains $n_{cp}=30a$ in a resolution of $\Delta\theta = 0.5a$. 
Since most secondary maxima are $<1\%$ 
we ignore them and therefore conclude on one global change point at $\hat\theta=1898$ in the interval $[1896,1900]$ 
of about $90\%$ confidence. Note that we interpret the splitting of the global maximum as an artefact from the high resolution of 
the numerical change point sampling $\Delta\theta = 0.5a$.\\
In conclusion, we are able to confirm previous investigation techniques and auxiliary reveal further information from 
the parameter space of the multidimensional posterior density of the applied LMM.


\section{Conclusions}
\label{conclusions}

\begin{table}
\caption{Estimated model of the annual Nile flux.}
\label{tab:2}   
\begin{tabular}{cll}
\hline
\hline\noalign{\smallskip}
parameter \hspace{2mm} & estimate \hspace{2mm} & confidence $\geq90\%$  \\
\noalign{\smallskip}\hline\noalign{\smallskip}
$\hat\theta$ & 1898 & [$\,$1895$\,$,$\,$1901$\,$] \\
$\hat\beta_{0}$ & 1.12 & [$\,$1.01$\,$,$\,$1.22$\,$] \\
$\hat\beta_{1}$ & -0.0013 & [$\,$-0.0082$\,$,$\,$0.0057$\,$] \\
$\hat\beta_{2}$ & 0.0006 & [$\,$-0.0011$\,$,$\,$0.0024$\,$] \\
$\hat\beta_{3}$ & 0.82 & [$\,$0.76$\,$,$\,$0.90$\,$] \\
$\hat\sigma$ & 0.124 & [$\,$0.094$\,$,$\,$0.160$\,$] \\
$\hat s_{1}$ & 0.0065 & [$\,$-0.0190$\,$,$\,$0.0450$\,$] \\
$\hat s_{2}$ & -0.0016 & [$\,$-0.0065$\,$,$\,$0.0855$\,$] \\
\noalign{\smallskip}\hline
\hline
\end{tabular}
\end{table}

We introduce a general method for the detection of trend changes in heteroscedastic time series by describing the 
observations as a linear mixed model. The change point is thereby considered as an isolated singularity in a regular 
background of a signal, assuming partial linear mean and deviation in the first order approach. 
By addressing the framework of linear mixed models we achieve to simplify the explicit computation of the marginal 
posterior distributions and thus reduce the computational time considerably. 
The formulation of the marginalized posterior densities of the model's parameters enables us to obtain inter alia 
the probability density of a change point given the data. Therefore the technique yields an insight in the 
parameter space of the underlying model, estimates these parameters and intrinsically provides a 
description of their confidence intervals. \\
We elaborate our technique for single change point models by infering on the relevant model parameters and discuss
the sensitivity of the singularity estimator with respect to data loss.
Additionally we present a kernel based approach to investigate more complex time series with multiple change points 
by localizing the posterior density and using the Bayes factor as a weighting function.\\
Moreover we apply our algorithm on the annual flow volume of the Nile River at Aswan from 1871 to 1970. We confirm 
a well-established transition in the year 1899 by the estimated change point at 1898 within the interval [$\,$1896$\,$,$\,$1900$\,$] of 
about $90\%$ confidence. We specify the underlying model and identify the mean as the statistical property undergoing the most 
significant transition.\\
We conclude by emphasizing that our algorithm depicts a powerfull tool to estimate the location of transitions in 
heteroscedastic time series and to infer on the underlying behavior in a partial linear approach, meanwhile reducing 
the computational time.

\section*{Acknowledgments}

We thank M.H. Trauth for fruitful discussions and gratefully acknowledge financial support by DFG 
(GRK Nadi and GRK 1364) and the University of Potsdam.

\end{document}